\documentstyle[12pt]{article}
\begin{document}
\def\strut{\rule[-.5cm]{0cm}{1cm}}
\def\dspace{\baselineskip = .30in}

\title{
\begin{flushright}
{\large\bf IFUP-TH 61/94}
\end{flushright}
\vspace{1.5cm}
\Large\bf IS THERE A DOMAIN WALL PROBLEM?}

\author{{\bf Gia  Dvali}\thanks{Permanent address: Institute of Physics,
Georgian Academy of Sciences,  \hspace{1cm}380077 Tbilisi, Georgia.
E-mail:
dvali@mvxpi1.difi.unipi.it }\\ Dipartimento di Fisica, Universita di Pisa
and INFN,\\ Sezione di Pisa I-56100 Pisa, Italy\\
\and
{\bf Goran Senjanovi{\'c}}\thanks{E-mail: goran@ictp.trieste.it}\\
International Center for Theoretical Physics\\
Trieste, Italy\\}

\date{ }
\maketitle

\begin{abstract} We show that
spontaneous breaking of discrete or continuous symmetries does
not necessarily imply either symmetry
restoration  or the thermal production of defects
 at high temperature (at least up to
$T \sim M_{Planck}$). This may imply that there is no domain wall
 problem. As an example we show how this applies to the Peccei-Quinn scenario.
\end{abstract}
\newpage

\dspace

\section{ Introduction}

We know from daily life that in the process of being heated physical systems
normally undergo phase transitions becoming less ordered. By analogy one
suspects the same of the
field theory systems with spontaneous symmetry breaking; at sufficiently
high temperature the order parameter or the vacuum expectation value
(VEV) of the scalar field should vanish, leading to symmetry restoration.
It turns out that this is precisely what happens in the case of a single
scalar field [1].If true in general, this would lead in many cases to
the
production of topological defects during the phase transition [2]. Some
of these defects such as domain walls,
 are disastrous for cosmology since in the context of the
standard big bang scenario they carry too much energy. This is
known as the domain wall problem [3].

Various remedies have been offered to this problem, the most celebrated
one being inflation.  However, this beautiful mechanism
would be of no help in many interesting cases when the temperature is
below the scale of inflation whose era is  expected to end at very
high scales (around $10^{16} GeV$ or so) [4].

 In the present paper we address the question whether the symmetry
nonrestoration can provide a solution to the domain wall problem.
The fact that symmetries may remain broken
at high $T$ was found long ago [5][6], however
 to our knowledge, the possible role of the symmetry nonrestoration
for the solution of the domain wall problem was in fact
never studied.

This has prompted us to reconsider this important issue of the high
temperature behavior of gauge theories.
 Much to our surprise, we find out that the nonrestoration of symmetry
at high $T$ seems to be a natural consequence in many minimal and
realistic models (in particular in the
theories of spontaneously broken P, CP and Peccei-Quinn symmetries).
We shall present these findings in a separate publication; here we show
how this happens in the invisible axion model.
Furthermore it turns out that the requirement that the dangerous domain walls
(or strings) are not produced thermally, at least up to temperatures of
order $M_P$, implies additional constraints on the parameters.

\dspace

\section{ High $T$ symmetry nonrestoration and topological defects}

Let us first recall briefly the essential features  of both symmetry
restoration and non-restoration mechanisms at high temperature.
The symmetry restoration at high $T$  can be illustrated on the
example of real scalar field $\phi$ with the Lagrangian

\begin{equation}
L(\phi) = {1\over 2}(\partial_{\mu} \phi)^2 -
{\lambda \over 4}(\phi^2 -\eta^2)^2
\end{equation}

This Lagrangian possesses a discrete $Z_2$ symmetry $\phi \rightarrow
-\phi$ spontaneously broken by $ \langle \phi \rangle = \pm \eta$.
The dominant
high temperature contribution to the effective potential for $T>>\eta$
is given by
\begin{equation}
\Delta V(T) = {\lambda\over 8} T^2\phi^2
\end{equation}
Since the boundedness of the potential demands $\lambda >0$,
for $T>>T_C=2\eta$, this leads inevitably to the restoration
of the $Z_2$ symmetry.

Now, according to the standard Kibble [2] scenario this fact
leads to the production of domain walls during the phase transition
when the universe cools down below $T_C$.

 However, the situation may change drastically in theories with more
fields. To see this, take a simple example of two real scalar
fields $\phi_1$ and $\phi_2$ with a potential

\begin{equation}
 V =  {\lambda_1 \over 4}(\phi_1^2 -\eta_1^2)^2
+ {\lambda_2 \over 4}(\phi_2^2 - \eta_2^2)^2 -
{\lambda \over 2}\phi_1^2\phi_2^2
\end{equation}
which has a $Z_2\otimes Z_2$ symmetry: $\phi_1\rightarrow -\phi_1,
\phi_2 \rightarrow \phi_2$ and vice versa. Now the boundedness
of the potential requires

\begin{equation}
\lambda_1>0,~~~~  \lambda_2>0,~~~~   \lambda_1\lambda_2>\lambda^2
\end{equation}
At high $T$ the potential receives the correction [5]

\begin{equation}
 \Delta V(T) = {1\over 24}[(3\lambda_1-\lambda)\phi_1^2 +
(3\lambda_2 -\lambda)\phi_2^2]T^2
\end{equation}

 One and $only$ one of these mass terms can be negative without
conflicting (4) meaning that one VEV can remain nonzero at high
$T$. Thus the symmetry is only partially restored to a single
$Z_2$.

\subsection*{A. Domain Walls}
Now it is time to ask: what about the domain wall problem in such theories?
For simplicity let us return to the case with two $Z_2$ symmetries.
First of all, the reader should not be confused by only partial
nonrestoration since in realistic cases
at least one of the fields transforms under a continuous gauge
$SU(2)\otimes U(1)$ symmetry. As a result one of the $Z_2$ factors
automatically becomes a part of the $SU(2)\otimes U(1)$ group and
obviously can not lead to the
domain wall formation. Thus, we always need to `nonrestore' only one
actual $Z_2$ at high $T$. Let us choose as such  the
$\phi_1 \rightarrow -\phi_1$
symmetry in our toy model, meaning that we will assume

\begin{equation}
3\lambda_1 - \lambda < 0
\end{equation}

Now at high $T$ the effective potential is minimized by
\begin{equation}
\langle \phi_1 \rangle = \pm T({\lambda -3 \lambda_1 \over
12 \lambda_1})^{1/2}
\label{vev}
\end{equation}
and the symmetry
is never restored. Although the order parameter (VEV of $\phi_1$) is growing
with temperature, the thermal production of domain walls will anyway take
place at some (very high) $T$. That is, if we start heating the
system in a homogeneous `initial condition', in which at $T = 0$
the vacuum is in say (+) phase everywhere, the thermal fluctuations will
finally destroy this picture by pulling  $\langle \phi_1 \rangle$ over the
potential barrier and therefore creating domains of new (-) phase separated
from the `old' ones by domain walls. However, such effect could only take
place at very high temperatures (in the very early universe).
  Production of a
spherically symmetric wall of radius $R$ is equivalent to the creation of a
vacuum bubble of the different phase.

The corresponding rate per unit time  and unit volume is given by [7]
\begin{equation}
\Gamma = T^4 \left( {S_3 \over 2 \pi T} \right)^{3/2}
e^{-{S_3 \over T}}
\label{rate}
\end{equation}
where  $S_3$
is the energy of the bubble (domain wall)
\begin{equation}
S_3 = 4 \pi R^2 \sigma
\label{energy}
\end{equation}
and $\sigma$ is the energy density per unit area [8] of a planar wall
\begin{equation}
\sigma = {4 \over 3} \sqrt{\lambda_1 \over 2} \langle \phi_1 \rangle^3
\label{energydensity}
\end{equation}

Now, a
sphericall wall of thickness larger than its own radius would simply
mean that no domain of the opposite phase is formed since the Higgs
field is in the false vacuum $everywhere$ inside the given sphere.
So, only walls whose size is bigger than their width should be considered
as being `formed'. The width of a domain wall is given by
\begin{equation}
\delta = \sqrt{  2\over \lambda_1} \langle \phi_1 \rangle^{-1} = 2
\sqrt{6 \over \lambda - 3\lambda_1} T^{-1} \label{width}
\end{equation}
and so we have $R > \delta$. If you assume for the moment $R\gg\delta$,
from (\ref{vev}),(\ref{energy}) you get
\begin{equation}
{S_3 \over T} \simeq {2 \pi \over 9\sqrt{6}} {(\lambda -
3\lambda_1)^{3/2} \over \lambda_1} R^2 T^2
\end{equation}
 and thus

\begin{equation}
{S_3 \over T} \gg  {16 \pi \over 3\sqrt{6}} {\sqrt{\lambda - 3\lambda_1}
\over \lambda_1}
\end{equation}
 One can see that, for
   $ \lambda_1$ small, the above suppression factor
 can be as small as one wishes and the numerical factors work in our favor.
Of course, as $R$ gets closer to $\delta$, the numerical factors become
less certain. However, the qualitative feature remains: for small
$\lambda_1$ one gets a large suppression.

  Of course, this analysis breaks down for temperatures close
to the Planck scale.
Since we really do not know what might happen at such high temperatures,
one may hope that the Planck scale corrections can modify
our effective potential in such a way that domain wall production even there
never happens. This is  a subject on which we can not speculate, but
the important
message from our analysis is that one can naturally raise the formation
temperature (say for the electroweak scale domain walls) by 16 to 17
 orders of
magnitude! Consequences are straightforward:
$a)$ either the walls are never formed or
$b)$ if formed, inflation, even if it takes place
at scales close to $M_P$, can cure any domain wall problem
including those
attributed to the electroweak symmetry breaking.

\subsection*{B. Strings}

In some cases, for example in the famous Peccei-Quinn scenario for the
solution of the strong CP problem, the domain wall problem is a consequence
of the existence of cosmic strings in the model.
 Thus in the context of symmetry nonrestoration the solution of
the axionic domain wall problem  reduces to the elimination of the string
producing phase transition. In complete analogy with the domain wall case
above, we can estimate the string production rate.  For this, assume that
in our toy model one of the Higgs fields (say $\phi_1$) is transforming
under  a global $U(1)$ symmetry. Since $U(1)$ is spontaneously broken, this
model admits a topologically stable global string solution [8].

 Now much in the way of  the $Z_2$ symmetry, the high $T$ correction
to the $\phi_1$-dependent part of the potential is
\begin{equation}
\Delta V(T,\phi_1) = {4 \lambda_1 -\lambda \over 24} T^2 |\phi_1|^2
\end{equation}

so that for $4 \lambda_1 -\lambda < 0$, the $U_1$ symmetry is never restored,
with
\begin{equation}
| \langle \phi_1(T) \rangle | = \left({\lambda - 4 \lambda_1
 \over 12 \lambda_1}\right)^{1/2} T
\end{equation}
at high $T$.

Again, the rate of the thermal production of a closed string with radius $R$
is given by (\ref{rate}), where $S_3 = 2 \pi R \mu$ and $\mu$ is the energy
 per unit length
\begin{equation}
\mu = \int \left[ |\nabla\phi_1|^2 + V(\phi_1) -
V(\langle \phi_1 \rangle)\right] d^2x
\label{energylenght}
\end{equation}

To estimate $\mu$ it is useful to separate the contributions inside
and outside of the core of the
string, ie to assume for the straight string (in cilindrical coordinates
$\rho$, $\theta$, $z$) \begin{equation}
\phi_1(x) = \langle \phi_1 \rangle  f(\rho) e^{i\theta}
\label{inside}
\end{equation}
where $f(\rho)$ is some (monotonic) function such that $f(0) = 0$ and
$f(\rho) = 1$ for any $\rho > \delta$. Here $\delta$ is the thickness of the
string and obviously we assume $R \gg \delta$. (In reality $f$ will
approach $1$ exponentially as one goes to infinity). Then the general
 form of $\mu$ is
\begin{equation}
\mu  =  2 \pi \langle \phi_1 \rangle^2 \left[ K(\delta) +  \ln{R\over
\delta} \right]
\label{mun}
\end{equation}
where $K$  comes from the core contribution
and depends on the explicit form of $f$. For any given ansatz $f$, the
thickness is determined through minimization of $\mu$ as
$\delta^{-1} = (dK / d\delta)$, giving $K$ a number of order one.
For example, for the linear dependence
inside the core $f=\rho / \delta$ we have $K = 3/2$ and
$\delta  = \sqrt{12 / \lambda_1} \langle \phi_1 \rangle^{-1} $ and thus from
$\mu
> 2\pi \langle \phi_1 \rangle^2$ we find the suppression factor for the
 string production
\begin{equation}
{S_3 \over T} >   4 \pi^2 \langle \phi_1 \rangle^2 {R \over T}
\end{equation}

which for $R\gg \delta$ gives
\begin{equation}
{S_3 \over T} \gg   4 \pi^2   {\sqrt{\lambda - 4 \lambda_1} \over \lambda_1}
\end{equation}

As in the case of walls, the thermal production rate for strings is
suppressed for small $\lambda_1$. Notice that the numerical factor in
front is already large. However, again for $R$ close to $\delta$ it
cannot be trusted.

 \section{Strong CP problem and domain walls}

 Here we wish to discuss how the possibility of symmetry nonrestoration may
solve
the infamous  domain wall problem of the Peccei-Quinn mechanism [9]. This
mechanism, commonly accepted as the solution to the strong CP problem, is
based on the concept of the continuous anomalous symmetry $U(1)_{PQ}$,
whose explicit breaking by instantons fixes $\bar\theta$ to be naturally
small. The instanton effects in the Higgs sector can be mimicked by the
effective phase dependent term in the potential [10]
\begin{equation}
\Delta V = \Lambda_{QCD}^4(1-cosN\theta)
\end{equation}
where $N$ is an integer and $a=\theta M_{PQ}$ is the axion field.
For example, in  the invisible axion extension [11] of the original
Peccei-Quinn model $N$ is the number of quark flavors. Thus, instantons
preserve a discrete subgroup of $U(1)_{PQ}$ characterized by
$\theta \rightarrow \theta + 2\pi/N$. The eventual spontaneous breaking
of this discrete symmetry leads to the formation of domain walls
[12].
The dynamics of the domain wall formation goes as follows [13].
At the scale $M_{PQ}$ when $U(1)_{PQ}$ is broken spontaneously the
network of global axionic cosmic strings is formed and $\theta$ winds
by $2\pi$ around each minimal string. Later on, at the temperature
$T\sim \Lambda_{QCD}$ the instanton effects are switched on and it
becomes energetically favorable to choose one out of the discrete
set of values $2\pi k/N$ ($k=1,2,..N$). Since $\Delta \theta =2\pi$
around the string this results in the formation of $N$ domain
walls attached to the string. The domain walls are topologically stable
and thus cosmologically troublesome for $N>1$.

 Note that a remarkably simple way out would be not to have strings
formed at all, which as we have  just seen could result  from the
nonexistence of the first phase transition. In such a case, above
$T\simeq \Lambda_{QCD}$, $\theta$ would be aligned having some typical
value $\theta_0$ which after the QCD phase transition would relax to
the nearest minimum (unless of course by some miracle $\theta_0$ would
turn out to lie at one of the local maxima). Now, the minimal realistic
PQ model is based on the introduction of a singlet field on top of
two Higgs doublets [11].
As the reader knows by now, it is perfectly natural to keep the VEV
of the singlet nonvanishing at high $T$, thus avoiding the formation
of cosmic strings and the subsequent troublesome domain walls.

Notice that the nonrestoration of $U(1)_{PQ}$ at high $T$ is impossible
without the singlet. Namely, with two doublets only one of the VEVs
can remain nonzero at high $T$ (since $U(1)_{PQ}$ forbids terms linear
in the fields) implying a necessary restoration of $U(1)_{PQ}$.
We find it remarkable that the singlet is essential for the solution
of the domain wall problem, since it is  needed for completely
different reasons, both phenomenological and astrophysical [10].

\section{Summary and Outlook}

As we have shown here, even in the minimal models there may be no
 domain wall problem. We must stress $may~not$,
since one does not know what can happen when one approaches the Planck
scale. The crucial point is that the usual scenario of Kibble may
not hold in general and whether or not there is a phase transition
depends on as of yet unknown parameters in the theory in question.

Furthermore, you can show that for a large range of parameters,
the thermal production of domain walls and strings never takes
place (except possibly for $T\simeq M_P$). This means that, even
in the case that these objects do get produced for $T$ near the
Planck scale, there is plenty of time for inflation to dilute the
density of their remnants. This provides a solution to the domain
wall problem even for the case of electroweak scale breaking of
the discrete symmetry in question. Finally, this would also imply
a way out of the infamous domain wall problem in the Peccei-Quinn
scenario with the invisible axion.

\section*{Acknowledgments}

We are grateful to Alejandra Melfo for discussions.

  \section*{References}

\begin{enumerate}

\item D.A.Kirzhnits and A.D.Linde, {\it Phys.Lett.}, {\bf B42} (1972) 471;

\item T.W.Kibble, {\it J.Phys.}, {\bf A9} (1976) 1987; {\it Phys.Rep.}
{\bf 67} (1980) 183;

\item Ya.B.Zeldovich, I.Yu.Kobzarev and L.B.Okun, {\it JETP}, {\bf 40}
(1974) 1;

\item A.D.Linde, {\it Particle Physics and Inflationary Cosmology}
(Harwood Academic, Switzerland, 1990); E.W.Kolb and M.S.Turner,
{\it The Early Universe} (Addison-Wesley, Reading, MA, 1990);

\item S.Weinberg, {\it Phys.Rev.}, {\bf D9}, (1974) 3357;

\item R.N.Mohapatra and G.Senjanovi{\'c}, {\it Phys.Rev.Lett},
{\bf 42}, (1979) 1651; {\it Phys.Rev.} {\bf D20}, (1979) 3390

\item A.D.Linde, {\it Phys.Lett.}, {\bf B100}, (1981) 37;
{\it Nucl. Phys.}, {\bf B216}, (1983) 421;

\item A.Vilenkin, {\it Phys. Rep.}, {\bf 121}, (1985) 263;

\item R.D.Peccei and H.R.Quinn, {\it Phys.Rev.} {\bf D16}
(1977) 1791;

\item J.E.Kim, {\it Phys.Rep.} {\bf 150} (1987) 1

\item M.A.Shifman, A.I.Vainshtein and V.I.Zakharov, {\it Nucl.Phys.}
{\bf B166} (1980) 493; M.Dine, W.Fischler and M.Srednicki, {\it Phys.Lett.}
{\bf B104} (1981) 199; A.P.Zhitnitskii, {\it Sov. J. Nucl.} {\bf 31}
(1980) 260;

\item P.Sikivie, {\it Phys.Rev.Lett} {\bf 48} (1982) 1156;

\item A.Vilenkin and A.Everett, {\it Phys.Rev.Lett.} {\bf 48}
(1982) 1867;

\end{enumerate}
\end{document}